\documentclass[hyper]{JHEP} 
\usepackage{graphicx,amssymb,bm,latexsym,amsmath}
\newcommand\fverb{\setbox\pippobox=\hbox\bgroup\verb}
\newcommand\fverbdo{\egroup\medskip\noindent%
            \fbox{\unhbox\pippobox}\ }

\newcommand\fverbit{\egroup\item[\fbox{\unhbox\pippobox}]}
\newbox\pippobox
\title{On the Rotating and Oscillating strings in $(AdS_3\times S^3)_{\varkappa}$}
\author{Aritra Banerjee\\
Department of Physics, Indian Institute of Technology Kharagpur,\\
Kharagpur-721 302, INDIA \\
Email: \email{aritra@phy.iitkgp.ernet.in}}
\author{Kamal L. Panigrahi\\
Department of Physics, Indian Institute of Technology Kharagpur,\\
Kharagpur-721 302, INDIA  and \\
Department of Physics, CERN Theory Division, CH-1211, Geneva 23,
Switzerland\\
Email: \email{panigrahi@phy.iitkgp.ernet.in}}
\abstract{We study rigidly rotating strings in the
$\varkappa$-deformed $AdS_3 \times S^3$ background. We find out
two classes of solutions corresponding to the giant magnon and
single spike solutions of the string rotating in two
$S^2_{\varkappa}$ subspace of rotations reduced along two
different isometries. We verify that the dispersion relations
reduce to the well known relation in the $\varkappa\rightarrow 0$
limit. We further study some oscillating string solutions in the
$S^3_{\varkappa}$ subspace. }

\begin{document}
\section{Introduction}
Integrability has played an important role in understanding the
spectrum of the superstring theory on $AdS_5\times
S^5$\cite{Beisert:2010jr}. This fact has been one of the key
ingredient in exploring the conjectured AdS/CFT duality
\cite{Maldacena:1997re} because of the fact that in planar limit
both sides of the duality, namely the gauge theory and string
theory, are more tractable \cite{Pohlmeyer:1975nb},
\cite{Minahan:2002rc}, \cite{Tseytlin:2004xa},
\cite{Hayashi:2007bq}, \cite{Okamura:2008jm}. This stems from the
basic fact that bosonic rigid spinning strings in the $AdS_5
\times S^5$ space-time are naturally described as periodic
solutions of the finite dimensional integrable system. It was
further noticed that the string theory is also integrable in the
semiclassical limit and the anomalous dimension of the ${\cal N} =
4$ Super Yang Mills (SYM) can be derived from the relation between
conserved charges of the rotating strings in  $AdS_5 \times S^5$.

There have been many ideas to introduce various classes of
integrable deformations to the string sigma model on $AdS_5\times
S^5$. Typically they are constructed by applying T-duality on any
given integrable model(e.g.
\cite{Lunin:2005jy},\cite{Frolov:2005ty},\cite{Frolov:2005dj},
\cite{Alday:2005ww},\cite{Ricci:2007eq},\cite{Beisert:2008iq}).
Contrary to the usual wisdom, however more recently, a novel
example of a one-parameter integrable deformation of the $AdS_5
\times S^5$ supercoset model was found in \cite{Delduc:2013qra},
following earlier proposals \footnote{Also in
\cite{Kawaguchi:2014qwa}, \cite{Matsumoto:2014nra} non-standard
q-deformed model based on the classical r-matrix satisfying
classical Yang-Baxter equations was explored.}. This model is
parameterized by a real deformed parameter $\eta\in[0,1)$ and
string tension $T \equiv g = \frac{\sqrt{\lambda}}{2\pi}$. This
model appears to be quite complicated and involved due to the
presence of fermionic degrees of freedom. As a first step to
understand the background better in \cite{Arutyunov:2013ega} a
Lagrangian corresponding to the bosonic degrees of freedom has
been studied in detail. Infact, the tree-level bosonic S-matrix
was also computed and quite successfully matched with the
semi-classical limit of the q-deformed S-matrix computed in
\cite{Beisert:2008tw}, \cite{Beisert:2011wq}. In a related
development, the bosonic subsectors of the q-deformed $AdS_5
\times S^5$ superstring action was studied and the classical
integrable structure of anisotropic Landau-Lifshitz sigma models
was derived by taking fast-moving limits in
\cite{Kameyama:2014bua}. Recently in \cite{Arutyunov:2014cda} the
bosonic spinning strings in the $\eta$-deformed $(AdS_5\times
S^5)$ have been studied, and shown that they are naturally
described by the periodic solution of a new finite dimensional
integrable system, which in turn could be viewed as a deformed
Neumann model. Restricting the motion of the string to the
deformed sphere, the Lax representation for the deformed Neumann
model has been presented. In \cite{Hoare:2014pna} various
deformation limits are discussed. In particular, in the so called
maximal deformation $(\eta = 1)$, the $AdS_5 \times S^5$ is transformed to
the T dual of the double wick rotated version of iteself, i.e.
$ dS_5 \times H^5$, where $dS_5$ is the five dimensional
de-Sitter and the $H^5$ is the five dimensional hyperboloid. The
corresponding worldsheet theory is non-unitary, nevertheless the
later is a solution to the type IIB supergravity equations of motion with a
imaginary five form flux. Infact more recently this problem of
non-unitarity was resolved in \cite{Arutyunov:2014cra} in a
background known as mirror space \cite{Arutyunov:2007tc} which is
formally related to $dS_5 \times H^5$ by a double $T$-duality. The
mirror sigma model, which is derived by taking a double Wick
rotation in the worldsheet theory of light cone $AdS_5\times S^5$,
inherits the symmetries, and the integrability of the light cone
$AdS_5 \times S^5$ sigma model, and is well-behaved. The
corresponding metric admits a lift to a full solution of the
standard type IIB supergravity as well. There exists also the
so called ``imaginary deformation" $\eta = i$, in which the total
10d metric transforms into a pp-wave like background having a
curved transverse part. In \cite{Hoare:2014pna}, using a related
parameter $\varkappa$ \footnote{$\varkappa = \frac{2\eta}{1-\eta^2}$
\cite{Arutyunov:2013ega}, where $\varkappa\in[0,\infty).$},
the $6d$ and $4d$ reductions of the total 10d metric have been performed and $(AdS_3 \times
S^3)_{\varkappa}$ and $(AdS_2\times S^2)_{\varkappa}$ backgrounds
have been proposed. The integrability of the $6d$ and $4d$ stems
from the fact that the original $10d$ spacetime is integrable.
Knowing the integrability of the sigma model in the limiting
background, it is interesting to explore the dual gauge theories
and the corresponding stringy states in the string theory side. It is
notable that the deformed 10d background breaks the
$SO(2,4)\times SO(6)$ symmetry of $AdS_5\times S^5$ into $[U(1)]^6$
, making the dual field theory description obscure. Of course the
role of this deformation parameter $\varkappa$ in the dual CFT is
not clear at all. In
order to know more about the boundary field theory, it is imperative
to investigate various rotating and pulsating strings in the
gravity side and then look for the relevant operators. In this
connection, in a subspace of the deformed $AdS_5\times S^5$, the
so called giant magnon solution was proposed and the magnon
excitation energy was computed in \cite{Arutynov:2014ota}. In the
deformed parameter $\varkappa \rightarrow 0$
limit, it reduces to the form of usual giant
magnon dispersion relation proposed by Hofman-Maldacena (HM)
\cite{Hofman:2006xt}. The HM giant magnon dispersion relation was
derived in the string theory side by looking at the rigidly
rotating strings in the $AdS_5 \times S^5$.

General rotating and pulsating string solutions in the
semiclassical limit have been very useful in understanding the
AdS/CFT like dualities in various backgrounds. The semiclassical
calculations in the string theory side has shown that the multi
spin rotating and pulsating string solutions beyond their BPS
limit with large charges are in perfect agreement with the ones
calculated in dual gauge theory. The pulsating strings in general
correspond to the highly excited sigma model operators. In this
connection a large number of rotating and pulsating string
solutions have been studied in various string theory backgrounds,
see for example, \cite{Kruczenski:2004wg}, \cite{Bobev:2005cz},
\cite{Ryang:2004tq},\cite{Smedback:1998yn},
\cite{Chen:2006gea},\cite{Kruczenski:2006pk},\cite{Ryang:2006yq},
\cite{Kluson:2007qu}, \cite{Ishizeki:2007we},
\cite{Bobev:2007bm},\cite{Lee:2008sk}, \cite{David:2008yk},
\cite{Lee:2008ui}. Motivated by the recent surge of interest in
the string spectrum of deformed $AdS_5 \times S^5$ background that
in general preserves the integrability, here we find a class of
rotating and pulsating strings in this background
\cite{Hoare:2014pna}. We solve for the most general ansatz for the
rotating strings in various subspaces of $\varkappa$-deformed
$AdS_3\times S^3$ and look for classical solutions. We construct
two classes of solutions corresponding to the giant magnon
\cite{Arutynov:2014ota},\cite{Khouchen:2014kaa} and single spike
solutions from the equations of motion of a fundamental string in
this deformed background. \footnote{Finite size giant magnon
solution in $\eta$ deformed $AdS_5\times S^5$ has been
investigated in \cite{Ahn:2014aqa}.} Single Spike strings are a
general class of solutions corresponding to the higher twist
operators in gauge theory \cite{Kruczenski:2004wg}. Furthermore it
was also shown that the giant magnon can be thought of as a
subclass of more general solutions in the short wavelength limit.
Infact in \cite{Ishizeki:2007we} it was found out that both the
giant magnon and single spike solutions of the string on the
sphere can be thought of as two limiting solutions.

Rest of the paper is organized as follows. In section-2, we will
study the rotating strings in the $\varkappa$-deformed
$AdS_3\times S^3$ background. As explained in \cite{Hoare:2014pna}
this background reduced from the $\varkappa$-deformed $AdS_5 \times
S^5$ \cite{Arutyunov:2013ega} does not contain a NS-NS B-field
even though the original space has it. We find two classes of
solutions corresponding to the known giant magnon and new single
spike solutions of the F-string equations of motion in two
different subspaces of the $\varkappa$-deformed $AdS_3\times S^3$.
We write the relevant dispersion relations and check that in
$\varkappa \rightarrow 0$ limit they do reduce to the known
dispersion relations. We also provide some additional comments here.
Section-3 is devoted to the study of string
solutions which is pulsating in $S^3_{\varkappa}$. In section-4 we
conclude with some remarks.

\section{Rigidly rotating strings in $\varkappa$-deformed $AdS_3\times S^3$}
We are interested in the deformed $AdS_3 \times S^3$ metric
proposed in \cite{Hoare:2014pna} (a consistent reduction from the
deformed $AdS_5\times S^5$ \cite{Arutyunov:2013ega}). It is given
by
\begin{eqnarray} ds^2 = - h(\rho) dt^2 + f(\rho) d\rho^2 + \rho^2
d\psi^2 + {\tilde h}(r) d\varphi^2 + \tilde{f} (r) dr^2 + r^2
d\phi^2 \ , \label{metric}
\end{eqnarray}
where
\begin{eqnarray}
h(\rho) = \frac{1+\rho^2}{1-\varkappa^2 \rho^2}, \>\>\> f(\rho) =
\frac{1}{(1+\rho^2)(1-\varkappa^2 \rho^2)} \nonumber \\
{\tilde h}(r) = \frac{1-r^2}{1+\varkappa^2 r^2}, \>\>\> {\tilde
f}(r) = \frac{1}{(1-r^2)(1+\varkappa^2 r^2)} \ ,
\end{eqnarray}
and the NS-NS two form B field vanishes. We are interested in
studying rigidly rotating and pulsating strings in this background
in two different planes of rotation when the motion is restricted
to $R_t \times S^3_{\varkappa}$ only which can be obtained by
putting $\rho = 0$.\footnote{this has been argued by solving the
equations of motion of a fundamental string in the given
background in \cite{Khouchen:2014kaa}.} The relevant metric is
written as
\begin{eqnarray}
ds^2 = -dt^2 + \frac{1-r^2}{1+\varkappa^2 r^2} d\varphi^2 +
\frac{1}{(1-r^2)(1+\varkappa^2 r^2)} dr^2 + r^2 d\phi^2 \ .
\label{metric1}
\end{eqnarray}

\subsection{String rotating in ($\theta$, $\varphi$) plane}
We start by putting $\phi = {\rm constant}$ and $r = \cos\theta$
in (\ref{metric1}) to get the 2d subspace which is just the
1-parameter $\varkappa$-deformed $S^2$ solution presented in
\cite{Hoare:2014pna} as
\begin{equation}
 ds^2= -dt^2 + \frac{1}{1+\varkappa^2\cos^2\theta} d\theta^2 + \frac{\sin^2\theta}
 {1+\varkappa^2\cos^2\theta} d\varphi^2
\end{equation}
Our starting point is the Polyakov action of the string in this
$R_t \times S_{\varkappa}$-deformed background
\begin{equation}
S = -\frac{\hat{T}}{2}\int d\sigma d\tau
[\sqrt{-\gamma}\gamma^{\alpha \beta}g_{MN}\partial_{\alpha} X^M
\partial_{\beta}X^N ]\ ,   \label{action}
\end{equation}
where $\gamma^{\alpha \beta}$ is the world-sheet metric and
$\hat{T} = T\sqrt{1+\varkappa^2}$ is the effective string
tension\cite{Arutyunov:2013ega}. Under conformal gauge (i.e.
$\sqrt{-\gamma}\gamma^{\alpha \beta}=\eta^{\alpha \beta}$) with
$\eta^{\tau \tau}=-1$, $\eta^{\sigma \sigma}=1$ and $\eta^{\tau
\sigma}=\eta^{\sigma \tau}=0$. Here we have to mention that
when the deformation parameter $\varkappa$ has
a general value, it is not known whether the total 10d
background satisfies full type IIB supergravity field equations.
Specifically the RR fluxes and the nontrivial dilaton are not exactly known
in this case. Indeed, it was shown in \cite{Hoare:2014pna} that the full
$(AdS_3\times S^3)_{\varkappa}$ admits a dilaton and RR 3-form fluxes
in the $\varkappa\rightarrow i$ limit.
But since we are using the conformal gauge, the
worldsheet scalar curvature $R^{(2)}= 0$, so that the dilaton
will not affect the worldsheet action in general.

Variation of the action (\ref{action}) with
respect to $X^M$ gives us the following equation of motion
\begin{eqnarray}
2\partial_{\alpha}(\eta^{\alpha \beta} \partial_{\beta}X^Ng_{KN})
&-& \eta^{\alpha \beta} \partial_{\alpha} X^M \partial_{\beta}
X^N\partial_K  g_{MN} =0 \ ,
\end{eqnarray}
and variation with respect to the metric gives the two Virasoro
constraints,
\begin{eqnarray}
g_{MN}(\partial_{\tau}X^M \partial_{\tau}X^N +
\partial_{\sigma}X^M \partial_{\sigma}X^N)&=&0 \ , \nonumber \\
g_{MN}(\partial_{\tau}X^M \partial_{\sigma}X^N)&=&0 \ .
\end{eqnarray}
To study rotating strings in this background we use the following
ansatz
\begin{equation}
 t = \mu\tau,\\\\ \ \theta=\theta(y),\\\\ \ \varphi = \omega(\tau + h(y)),\\\ \
 y=\sigma-v\tau \ . \label{ansatz}
\end{equation}
The equation of motion for $\varphi$ gives
\begin{equation}
 \partial_y h = \frac{1}{1 - v^2}[\frac{A_1(1+\varkappa^2\cos^2\theta)}{\sin^2\theta} - v].
\end{equation}
$A_1$ here is an integration constant. Putting the above equation
into the equation of motion for $\theta$ we get
\begin{eqnarray}
  (1-v^2)^2(\frac{\partial\theta}{\partial y})^2  = &-&  \omega^2[\frac{A_1^2(1+\varkappa^2\cos^2\theta)^2}{\sin^2\theta}+
  \sin^2\theta] \nonumber\\ &+& A_2(1+\varkappa^2\cos^2\theta) \ .
\end{eqnarray}
Since we are interested in the infinite $J$ magnon, we put the
boundary condition when $\theta=\theta_{max} = \pi/2$ \ \ ,
$\frac{\partial\theta}{\partial y}\rightarrow 0$. This means $A_2
= \omega^2(A_1^2+1)$. Now putting these values back into the
$\theta$ equation of motion, we recover
\begin{equation}
 (1-v^2)^2(\frac{\partial\theta}{\partial y})^2= \omega^2(1+\varkappa^2)
 \cot^2\theta[\sin^2\theta(1+A_1^2\varkappa^2)-A_1^2(1+\varkappa^2)]
 \ ,
\end{equation}
which leads to the equation
\begin{equation}
 \frac{\partial\theta}{\partial y} = \frac{\omega\sqrt{(1+\varkappa^2)(1+A_1^2\varkappa^2)}
 \cot\theta}{(1-v^2)}\sqrt{\sin^2\theta-\sin^2\theta_{0}} \ ,
\end{equation}
where
\begin{equation}
\sin^2\theta_{0} = \frac{A_1^2(1+\varkappa^2)}{1+A_1^2\varkappa^2}
\ . \label{sin}
\end{equation}
Again subtracting the two Virasoro constraints we get a relation
between the constants
\begin{equation}
 \mu^2 - \frac{\omega^2 A_1}{v} = 0 \ . \label{relation}
\end{equation}
Also by equating the $\theta$ equation with the first Virasoro
constraint we get the equation for $A_1$ as
\begin{equation}
 A_1^2 - A_1\frac{1+v^2}{v} +1 = 0,
\end{equation}
the solutions of which gives the values of $A_1$ consistent with
the Virasoro constraints. We can see the roots of the above
equation correspond to two different limiting solutions which we
identify as
\begin{eqnarray}
 A_1 &=& v ~~~~~ {\rm magnon~case} \nonumber\\ &=& \frac{1}{v} ~~~~~ {\rm single~spike~case} \label{cases}
\end{eqnarray}
Now the symmetry of the background gives rise to the following
conserved charges
\begin{eqnarray}
E &=& -\int\frac{\partial \mathcal{L}}{\partial \dot{t}}d\sigma
\nonumber \\ &=&
\frac{\hat{T}}{\sqrt{(1+\varkappa^2)(1+A_1^2\varkappa^2)}}
\frac{\mu (1-v^2)}{\omega}
\int\frac{\sin\theta~d\theta}{\cos\theta\sqrt{(\sin^2\theta-\sin^2\theta_0)}} \, \nonumber\\
J_{\varphi}&=& \int\frac{\partial \mathcal{L}}{\partial\dot{\varphi}} d\sigma = \hat{T}\int\dot{\varphi}~g_{\varphi\varphi}~d\sigma \nonumber\\
&=&
\frac{\hat{T}}{\sqrt{(1+\varkappa^2)(1+A_1^2\varkappa^2)}}(1-vA_1)\int\frac{\sin\theta~d\theta}{\cos\theta\sqrt{(\sin^2\theta-\sin^2\theta_0)}}
\nonumber\\
&-& \hat{T}\sqrt{\frac{(1+\varkappa^2)}{(1+A_1^2\varkappa^2)}}
\int\frac{\sin\theta\cos\theta~d\theta}{(1+\varkappa^2\cos^2\theta)\sqrt{(\sin^2\theta-\sin^2\theta_0)}}.
\end{eqnarray}
The angle deficit in this case can be defined by
\begin{eqnarray}
   \Delta\varphi &=& \omega\int\frac{\partial h}{\partial y}dy \nonumber\\
   &=& \frac{1}{\sqrt{(1+\varkappa^2)(1+A_1^2\varkappa^2)}}\int(\frac{A_1(1+\varkappa^2\cos^2\theta)}{\sin^2\theta}-v)
   \frac{\sin\theta~d\theta}{\cos\theta\sqrt{(\sin^2\theta-\sin^2\theta_0)}} \
   . \nonumber \\
\end{eqnarray}
Now we wish to find the relation between the above Noether charges
in the limits mentioned in (\ref{cases}) using some particular
regularization scheme to avoid divergences.
\subsubsection{Giant magnon solution}
Using the limit $A_1 = v$ we evaluate the charges as explained
above. The expression for energy
\begin{equation}
 E = 2\frac{\hat{T}}{\sqrt{(1+\varkappa^2)(1+v^2\varkappa^2)}} \frac{\mu (1-v^2)}{\omega}
\int_{\theta_0}^{\pi/2}\frac{\sin\theta~d\theta}{\cos\theta\sqrt{(\sin^2\theta-\sin^2\theta_0)}}
\end{equation}
diverges in the upper limit. While the angular momenta looks like
\begin{eqnarray}
 J_{\varphi}&=& \frac{2\hat{T}}{\sqrt{(1+\varkappa^2)(1+v^2\varkappa^2)}}(1-v^2)\int_{\theta_0}^{\pi/2}\frac{\sin\theta~d\theta}{\cos\theta\sqrt{(\sin^2\theta-\sin^2\theta_0)}}
\nonumber\\
&-& 2\hat{T}\sqrt{\frac{(1+\varkappa^2)}{(1+v^2\varkappa^2)}}
\int_{\theta_0}^{\pi/2}\frac{\sin\theta\cos\theta~d\theta}{(1+\varkappa^2\cos^2\theta)\sqrt{(\sin^2\theta-\sin^2\theta_0)}},
\end{eqnarray}
which is also divergent evidently due to the first integral.
However the difference $\frac{\omega E}{\mu}-J_{\varphi}$ remains
finite. This can be easily seen by writing explicitly,
\begin{eqnarray}
 \tilde{E}-J_{\varphi} &=& 2\hat{T}\sqrt{\frac{(1+\varkappa^2)}{(1+v^2\varkappa^2)}}
\int_{\theta_0}^{\pi/2}\frac{\sin\theta\cos\theta~d\theta}{(1+\varkappa^2\cos^2\theta)\sqrt{(\sin^2\theta-\sin^2\theta_0)}}  \nonumber\\
&=&
\frac{2\hat{T}}{\varkappa}\sqrt{\frac{(1+\varkappa^2)}{(1+v^2\varkappa^2)
(1+\varkappa^2\cos^2\theta_0)}}\tanh^{-1}(\frac{\varkappa\cos\theta_0}{\sqrt{1+\varkappa^2\cos^2\theta_0}}).
\label{magnon1}
\end{eqnarray}
Here $\tilde{E}= \frac{\omega E}{\mu}$ is the re-scaled
energy\footnote{Here one can note that from (\ref{relation}) the
scaling factor $\frac{\omega}{\mu}$ can be safely put to be 1 in
this case.}. We can also evaluate the deficit angle, which is a
finite quantity in this limit, as
\begin{eqnarray}
  \Delta\varphi &=&  \frac{2v\sqrt{1+\varkappa^2}}{\sqrt{1+v^2\varkappa^2}}\int_{\theta_0}^{\pi/2}
   \frac{\cos\theta~d\theta}{\sin\theta\sqrt{(\sin^2\theta-\sin^2\theta_0)}} \nonumber\\
   &=& \frac{2v\sqrt{1+\varkappa^2}}{\sqrt{1+v^2\varkappa^2}} \frac{\cos^{-1}(\sin\theta_0)}{\sin\theta_0} \nonumber\\
   &=& 2\cos^{-1}(\sin\theta_0).
\end{eqnarray}
Here, we have used the value of $\sin\theta_0$ from (\ref{sin}).
So using the above expression, we can readily see that the finite
difference (\ref{magnon1}) takes the form
\begin{equation}
 \tilde{E}-J_{\varphi} = \frac{2\hat{T}}{\varkappa}\tanh^{-1}
 \left(\frac{\varkappa|\sin(\frac{\Delta\varphi}{2})|}
 {\sqrt{1+\varkappa^2\sin^2(\frac{\Delta\varphi}{2})}}\right) \ , \label{magnon2}
\end{equation}
Which is the giant magnon dispersion relation presented in
\cite{Khouchen:2014kaa}. It can be shown that under a
$\varkappa\rightarrow 0$ limit the above expression reduces to the
form of HM giant magnon on $\mathbf{R}\times S^2$
\begin{equation}
 \lim_{\varkappa\rightarrow 0} ~~ \tilde{E}-J_{\varphi} = 2T\sin(\frac{\Delta\varphi}{2})= \frac{\sqrt{\lambda}}{\pi}\sin(\frac{\Delta\varphi}{2}).
\end{equation}
Further it has also been noted in \cite{Khouchen:2014kaa} that the
giant magnon dispersion relation (\ref{magnon2}) can be shown to
agree with the magnon excitation energy calculated in
\cite{Arutynov:2014ota}, by using the relation between $\eta$ and
$\varkappa$ defined in \cite{Arutyunov:2013ega}.
\subsubsection{Single spike solution}
In the opposite limit solution we put $A_1 = \frac{1}{v}$, so the
expression for energy remains the same
\begin{equation}
 E = \frac{2\hat{T}}{\sqrt{(1+\varkappa^2)(1+\frac{\varkappa^2}{v^2})}} \frac{\mu (1-v^2)}{\omega}
\int^{\theta_0}_{\pi/2}\frac{\sin\theta~d\theta}{\cos\theta\sqrt{(\sin^2\theta-\sin^2\theta_0)}}.
\end{equation}
The angular momentum $J_{\varphi}$ now becomes finite as
\begin{eqnarray}
 J_{\varphi}&=& -2\hat{T}\sqrt{\frac{(1+\varkappa^2)}{(1+\frac{\varkappa^2}{v^2})}}
\int^{\theta_0}_{\pi/2}\frac{\sin\theta\cos\theta~d\theta}{(1+\varkappa^2\cos^2\theta)\sqrt{(\sin^2\theta-\sin^2\theta_0)}}\nonumber\\
&=&
\frac{2\hat{T}}{\varkappa}\tanh^{-1}(\frac{\varkappa\cos\theta_0}{\sqrt{1+\varkappa^2\cos^2\theta_0}}).
\end{eqnarray}
We can show that in the required limit the above reduces to the
known value as
\begin{equation}
 \lim_{\varkappa \rightarrow 0}~~J_{\varphi} = 2T\cos\theta_0.
\end{equation}
In this case we can also write the expression for deficit angle as
\begin{eqnarray}
 \Delta\varphi &=& \frac{2}{\sqrt{(1+\varkappa^2)(1+\frac{\varkappa^2}{v^2})}}
 \Big[(\frac{1}{v}-v)\int_{\pi/2}^{\theta_0}\frac{\sin\theta~d\theta}{\cos\theta
 \sqrt{(\sin^2\theta-\sin^2\theta_0)}}
 \nonumber\\
 &+&\frac{1+\varkappa^2}{v}\int_{\pi/2}^{\theta_2}\frac{\cos\theta~d\theta}
 {\sin\theta\sqrt{(\sin^2\theta-\sin^2\theta_0)}}\Big] \ .
\end{eqnarray}
This is now a divergent quantity, but we can subtract the
divergent integral by combining it with $E$ as follows
\begin{eqnarray}
 \bar{E}-\hat{T}\Delta\varphi &=& \frac{\omega E}{\mu v} - \hat{T}\Delta\varphi \nonumber\\
 &=& -\frac{2\hat{T}}{v}\frac{\sqrt{1+\varkappa^2}}{\sqrt{1+\frac{\varkappa^2}{v^2}}}\int_{\pi/2}^{\theta_0}
   \frac{\cos\theta~d\theta}{\sin\theta\sqrt{(\sin^2\theta-\sin^2\theta_0)}} \nonumber\\
   &=& \frac{2\hat{T}}{v}\frac{\sqrt{1+\varkappa^2}}{\sqrt{1+\frac{\varkappa^2}{v^2}}} \frac{\cos^{-1}(\sin\theta_0)}{\sin\theta_0} \nonumber\\
   &=& 2\hat{T}(\frac{\pi}{2}-\theta_0), \label{spike1}
\end{eqnarray}
which is analogous to the usual spike height relation, where
$\bar{\theta}=(\frac{\pi}{2}-\theta_0)$ denotes the height of the
spike.\footnote{Again we note that the scaling factor
$\frac{\omega}{\mu v}$ can be put to be unity from
(\ref{relation}) in this case.} This relation evidently reduces to
the usual spike height relation \cite{Ishizeki:2007we} in the
$\varkappa \rightarrow 0$ limit.

Before we close this section, let us make some remarks about the
fate of our solution in the two different limits of the background
discussed in \cite{Hoare:2014pna} i.e. i) $\varkappa=i$ and ii)
$\varkappa=\infty$. As explained in \cite{Hoare:2014pna}, in the
first limit, together with scaling of the coordinates, transforms
the background to a pp-wave geometry with the transverse part
having a non zero curvature. However the limit has to be taken
cautiously, because naively taking $\varkappa\rightarrow i$ will
make the effective string tension $\hat{T}$ vanish. For this
purpose we can follow the recipe pointed out in
\cite{Hoare:2014pna}  and take the limit
 in non trivial way as
\begin{equation}
\varkappa^2 = -1 + s\epsilon^2 ,  \,\,\, t  = \frac{x^+}{\epsilon}
- \epsilon x^- , \, \, \, \varphi  = \frac{x^+}{\epsilon} +
\epsilon x^-
\end{equation}
where $\epsilon\rightarrow 0$ and $s$ can be put to 1 after taking
the limit without loss of generality. For the simple deformed
$(\mathbf{R}\times S^2)$ case ($(\theta, \varphi)$ plane) we can
find the pp-wave type metric as
\begin{equation}
 ds^2 = 4 dx_+ dx_- - \sinh^2\beta (dx_+)^2 + d\beta^2 ,
\end{equation}
where we have used the transformation $r = \tanh\beta$. Evidently
when $\sinh\beta = \beta$, this reduces to the usual pp-wave
metric for $AdS$ backgrounds, which have been extensively studied.
However, studying string solutions in this type of pp-wave
backgrounds with general rotating string ansatz as proposed in
(\ref{ansatz}) may not be useful. However, one can study simpler
solutions like the ``straight strings" in the AdS-pp wave
background as studied in, for example, \cite{Ishizeki:2008tx}.

On the other hand $\varkappa=\infty$ limit relates the
$AdS_5\times S^5$ metric to the T-dual of double wick rotated
version of it, i.e. $dS_5\times H^5$. As mentioned earlier, this
background is a solution of IIB supergravity equations supported
by imaginary 5-form flux. Again, naively taking the $\varkappa
\rightarrow\infty$ limit on the dispersion relations will not be
meaningful. To proceed further, we start with the limiting
background derived from $AdS_2\times S^2$\cite{Hoare:2014pna}
\begin{equation}
 \varkappa^2 ds^2 = - \frac{d\widetilde{\rho}^2}{1+\widetilde{\rho}^2} + (1+\widetilde{\rho}^2)dt^2 +
 \frac{d\widetilde{r}^2}{\widetilde{r}^2 - 1} + (\widetilde{r}^2 - 1)d\varphi^2,
\end{equation}
where $\widetilde{r}= \frac{1}{r}$ and $\widetilde{\rho}=
\frac{1}{\rho}$. This geometry corresponds to that of $dS_2\times
H^2$ without any need of T-duality, albeit upto an overall
$\varkappa^2$ factor. Here, $\widetilde{\rho}$ acts like the new
time coordinate. String solutions in this kind of geometry has to
be understood in a better way.

\subsection{String rotating in ($\theta$, $\phi$) plane}
We start with putting $r = \sin\theta$ in the metric
(\ref{metric}) and using the general ansatz
\begin{equation}
 t = \mu\tau,\\\ \ \theta=\theta(y),\\\ \ \phi = \omega(\tau + g(y)),\\\ \ \varphi=constant, \\\ \  y=\sigma-v\tau
\end{equation}
The choice can be justified by the equation of motion for
$\varphi$
\begin{equation}
 \partial_{\alpha}[g_{\varphi\varphi}~\eta^{\alpha\beta}\partial_{\beta}\varphi] =
 0 \ ,
\end{equation}
which is satisfied by a constant value of $\varphi$. For this
geometry the string equation of motion are written as
\begin{equation}
 \partial_y g = \frac{1}{1 - v^2}\left[\frac{C_1}{\sin^2\theta} -
 v\right] \ .
\end{equation}
Here $C_1$ is the integration constant. Using these two above
equations and putting them into the $\theta$ equation of motion we
get,
\begin{eqnarray}
  (1-v^2)^2\left(\frac{\partial\theta}{\partial y}\right)^2  = &-&
  \omega^2\left[\frac{C_1^2(1+\varkappa^2\sin^2\theta)}{\sin^2\theta}+
  \sin^2\theta(1+\varkappa^2\sin^2\theta)\right] \nonumber\\ &+& C_2(1+\varkappa^2\sin^2\theta)
\end{eqnarray}
Again we would like to investigate the infinite $J$ string states, so we can put the
boundary condition when $\theta=\theta_{max} = \pi/2$ \ \
$\frac{\partial\theta}{\partial y}\rightarrow 0$. This means $C_2
= \omega^2(C_1^2+1)$. Now putting these values into the $\theta$
equation of motion, we recover
\begin{equation}
 (1-v^2)^2\left(\frac{\partial\theta}{\partial y}\right)^2= \cot^2\theta[(\omega^2 -\omega^2 C_2^2 \varkappa^2)\sin^2\theta
 + \omega^2 \varkappa^2 \sin^4\theta - \omega^2 C_1^2] \ ,
\end{equation}
which leads to the equation
\begin{equation}
 \frac{\partial\theta}{\partial y} = \frac{\omega\varkappa\cot\theta}{(1-v^2)}
 \sqrt{[(\sin^2\theta-\sin^2\theta_1)(\sin^2\theta-\sin^2\theta_2)]}
 \ .
\end{equation}
Here, we can note that the roots are $\sin^2\theta_1 =
-\frac{1}{\varkappa^2} < 0$ and $\sin^2\theta_2 = C_1^2 >  0$. In
this case the existence of a negative root can argued as in
\cite{Bykov:2008bj} for a large $J$ expansion. Since we have
already put one of the roots to be $\pi/2$ we would expect the
positive solution $\sin^2\theta_2\in(0,1)$ which will be justified
by the string state solutions in question. The Virasoro
constraints on the other hand will again lead to two limiting
values of $C_1$ as before, which will lead to the two independent
solutions, namely giant magnon and single spiky string solution.
Now looking at the isometries of the metric (\ref{metric1}) we can
write the conserved charges in this background as follows
\begin{eqnarray}
E &=& -\int\frac{\partial \mathcal{L}}{\partial \dot{t}}d\sigma
\nonumber \\ &=&
\frac{\hat{T}}{\varkappa}\frac{\mu(1-v^2)}{\omega}
\int\frac{\sin\theta~d\theta}{\cos\theta\sqrt{(\sin^2\theta-\sin^2\theta_1)
(\sin^2\theta-\sin^2\theta_2)}} \ , \nonumber\\
J_{\phi}&=& \int\frac{\partial \mathcal{L}}{\partial\dot{\phi}} d\sigma
= \hat{T}\int\dot{\phi}\sin^2\theta~d\sigma \ , \nonumber\\
&=&
\frac{\hat{T}}{\varkappa}[(1-vC_1)\int\frac{\sin\theta~d\theta}
{\cos\theta\sqrt{(\sin^2\theta-\sin^2\theta_1)(\sin^2\theta-\sin^2\theta_2)}}
\ ,
\nonumber\\
&-&
\int\frac{\sin\theta\cos\theta~d\theta}{\sqrt{(\sin^2\theta-\sin^2\theta_1)
(\sin^2\theta-\sin^2\theta_2)}}].
\end{eqnarray}
We can also define the angle deficit as
\begin{eqnarray}
   \Delta\phi &=& \omega\int\frac{\partial g}{\partial y}dy \nonumber\\
   &=& \frac{1}{\varkappa}\int\left(\frac{C_1}{\sin^2\theta}-v\right)
   \frac{\sin\theta~d\theta}{\cos\theta\sqrt{(\sin^2\theta-\sin^2\theta_1)(\sin^2\theta-\sin^2\theta_2)}}.
\end{eqnarray}
 Using these conserved quantities we can investigate the relationship between them for the two limits
 mentioned before.
\subsubsection{Giant magnon solution}
Using $C_1 = v$, we evaluate the conserved charges using the
integrals mentioned in the appendix. The expression for energy
\begin{equation}
 \tilde{E} = E\frac{\omega}{\mu} = \frac{2\hat{T}}{\varkappa}(1-v^2)
\int_{\theta_2}^{\pi/2}\frac{\sin\theta~d\theta}{\cos\theta\sqrt{(\sin^2\theta-\sin^2\theta_1)
(\sin^2\theta-\sin^2\theta_2)}},
\end{equation}
which diverges. Also in this limit the angular momenta
\begin{eqnarray}
 J_{\phi} &=& \frac{2\hat{T}}{\varkappa}[(1-v^2)\int_{\theta_2}^{\pi/2}
 \frac{\sin\theta~d\theta}{\cos\theta\sqrt{(\sin^2\theta-\sin^2\theta_1)(\sin^2\theta-\sin^2\theta_2)}}
 \nonumber\\
&-&
\int_{\theta_2}^{\pi/2}\frac{\sin\theta\cos\theta~d\theta}{\sqrt{(\sin^2\theta-\sin^2\theta_1)
(\sin^2\theta-\sin^2\theta_2)}}]
\end{eqnarray}
diverge due to the first integral. However the difference
$\tilde{E} - J_{\phi}$ remains finite and can be evaluated to be
\begin{equation}
 \tilde{E} - J_{\phi} = \frac{2\hat{T}}{\varkappa}\tanh^{-1}(\frac{\cos\theta_2}{\cos\theta_1}).
\end{equation}
However the angular deficit is finite in this case and can be
written as
\begin{eqnarray}
 \Delta\phi &=& \frac{2v}{\varkappa}\int_{\theta_2}^{\pi/2}
   \frac{\cos\theta~d\theta}{\sin\theta\sqrt{(\sin^2\theta-\sin^2\theta_1)(\sin^2\theta-\sin^2\theta_2)}}
   \nonumber\\
   &=& \frac{2v}{\varkappa\sin\theta_1\sin\theta_2}\tanh^{-1}(\frac{\sin\theta_1\cos\theta_2}
   {\cos\theta_1\sin\theta_2}).
\end{eqnarray}
Putting the values of $\sin\theta_1$ and $\sin\theta_2$ into the
above equation, we can see it translates to
\begin{equation}
 \sin\left(\frac{\Delta\phi}{2}\right) = \frac{\sqrt{1-v^2}}{\sqrt{1+v^2\varkappa^2}}.
\end{equation}
Now it can be easily proved that the conserved charges in this
case obey the dispersion relation as
\begin{equation}
 \tilde{E} - J_{\phi} = \frac{2\hat{T}}{\varkappa}\tanh^{-1}
 \left(\frac{\varkappa|\sin(\frac{\Delta\phi}{2})|}{\sqrt{1+\varkappa^2\sin(\frac{\Delta\phi}{2})}}\right), \label{magnon1}
\end{equation}
which is the same dispersion relation as mentioned in
\cite{Khouchen:2014kaa}. We can take a $\varkappa \rightarrow 0$ limit to
get
\begin{equation}
 \tilde{E} - J_{\phi} = 2T\sin(\frac{\Delta\phi}{2})= \frac{\sqrt{\lambda}}{\pi}\sin(\frac{\Delta\phi}{2}),
\end{equation}
which is indeed the giant magnon dispersion relation in
$\mathbf{R}\times S^2$ as mentioned in \cite{Hofman:2006xt}.

\subsubsection{Single spike solution}
Using the other viable limit $C_1 = \frac{1}{v}$, we again
evaluate the conserved charges as before. Here also the
$\tilde{E}$ remains divergent as
\begin{equation}
 \tilde{E} = \frac{2\hat{T}}{\varkappa}(1-v^2)
\int_{\pi/2}^{\theta_2}\frac{\sin\theta~d\theta}{\cos\theta\sqrt{(\sin^2\theta-\sin^2\theta_1)(\sin^2\theta-\sin^2\theta_2)}}.
\end{equation}
 While $J_{\phi}$ in this case is finite as
 \begin{eqnarray}
  J_{\phi} &=& -\frac{2\hat{T}}{\varkappa}
  \int_{\pi/2}^{\theta_2}\frac{\sin\theta\cos\theta~d\theta}{\sqrt{(\sin^2\theta-\sin^2\theta_1)(\sin^2\theta-\sin^2\theta_2)}} \nonumber\\
  &=& \frac{2\hat{T}}{\varkappa}\tanh^{-1}(\frac{\cos\theta_2}{\cos\theta_1}). \label{J}
 \end{eqnarray}
On the other hand the angle deficit
\begin{eqnarray}
 \Delta\phi &=& \frac{2}{\varkappa}
 [(\frac{1}{v}-v)\int_{\pi/2}^{\theta_2}\frac{\sin\theta~d\theta}{\cos\theta\sqrt{(\sin^2\theta-\sin^2\theta_1)(\sin^2\theta-\sin^2\theta_2)}}
 \nonumber\\
 &+&\frac{1}{v}\int_{\pi/2}^{\theta_2}\frac{\cos\theta~d\theta}{\sin\theta\sqrt{(\sin^2\theta-\sin^2\theta_1)(\sin^2\theta-\sin^2\theta_2)}}]
\end{eqnarray}
diverges due to the first integral present here. Although it is
easy to see that by combining, we get
\begin{eqnarray}
 \bar{E}-\hat{T}\Delta\phi &=& E\frac{\omega}{\mu v} - \hat{T}\Delta\phi \nonumber\\
 &=& -\frac{2\tilde{T}}{v\varkappa}\int_{\pi/2}^{\theta_2}
 \frac{\cos\theta~d\theta}{\sin\theta\sqrt{(\sin^2\theta-\sin^2\theta_1)(\sin^2\theta-\sin^2\theta_2)}} \nonumber\\
 &=& \frac{2\tilde{T}}{v\varkappa\sin\theta_1\sin\theta_2}\tanh^{-1}(\frac{\sin\theta_1\cos\theta_2}{\cos\theta_1\sin\theta_2}).
\end{eqnarray}
As we can see the spike height (and hence the energy) in this case
is modified non trivially in contrast to the previous case
(\ref{spike1}). Putting all the values and doing little algebra we
can see that under the limit $\varkappa \rightarrow 0$ this
relation again becomes
\begin{eqnarray}
 \bar{E}-T\Delta\phi &=& 2T \tan^{-1}(\frac{\cos\theta_2}{\sin\theta_2}) \nonumber\\
 &=& 2T(\frac{\pi}{2}-\theta_2), \label{spike2}
\end{eqnarray}
which is the usual spike-height relation as mentioned in
\cite{Ishizeki:2007we}. Even it can also be proved that under $\varkappa
\rightarrow 0$ limit the expression for angular momentum becomes
\begin{equation}
 J_{\phi} = 2T\cos\theta_2,
\end{equation}
as we get usually for the $\mathbf{R}\times S^2$ case. Now, few
comments are in order about the solutions derived above, mainly in
the light of the recent work \cite{Hoare:2014pna}. The background
(\ref{metric1}) has an unique property in the sense the $\phi$ and
$\varphi$ are related by a discrete $Z_2$ symmetry
\cite{Hoare:2014pna} which is manifested as
\begin{equation}
 \phi\rightarrow\varphi ; \,\,\, r \rightarrow \sqrt{\frac{1-r^2}{1+\varkappa^2 r^2}}.
\end{equation}
Surely, We can see that these symmetries are quite evident by the form of equations (\ref{magnon2})
and (\ref{magnon1}) as these two giant magnon dispersion relations are simply connected by the change
$\phi\rightarrow\varphi$. For the single spike solutions, from equations (\ref{spike1}) and (\ref{spike2}), it can be
seen that the height of the spike solutions depend on the two angles $\theta_0$ and $\theta_2$. Further, keeping in
mind that we have put $r = \sin\theta$ for the solutions in $(\theta,\phi)$ plane and $r = \cos\theta$ for
the solutions in $(\theta,\varphi)$ plane, we can relate the angles $\theta_0$ and $\theta_2$ as
\begin{equation}
 r_2 = \sin\theta_2 \rightarrow r_0=\cos\theta_0 = \sqrt{\frac{1-r_2^2}{1+\varkappa^2 r_2^2}},
\end{equation}
since $\sin\theta_2 = \frac{1}{v}$ and $\cos\theta_0 = \sqrt{\frac{v^2- 1}{v^2+\varkappa^2}}$ with
the appropriate choice of constants for the spiky string cases as expalined in the previous subsections.
This, coupled with $\phi\rightarrow\varphi$ relates (\ref{spike1}) and
(\ref{spike2}) explicitly. These symmetries have been used extensively in \cite{Hoare:2014pna} to relate
the $\varkappa$ deformed model to other classically integrable models.

\section{Pulsating strings in deformed $\mathbf{R}\times S^3$}
In this section, we
wish to study a class of strings pulsating in full deformed $S^3$ with an extra
angular momentum. Beginning with the metric (\ref{metric}) and
putting $r = \cos\theta$ we get the full deformed
$\mathbf{R}\times S^3$ metric as
\begin{equation}
 ds^2= -dt^2 + \frac{1}{1+\varkappa^2\cos^2\theta} d\theta^2 + \frac{\sin^2\theta}{1+\varkappa^2\cos^2\theta} d\varphi^2 + \cos^2\theta d\phi^2.
\end{equation}
The Polyakov action for the string in the above background is
given by
\begin{equation}
S = \frac{\hat{T}}{2}\int d\tau d\sigma\Big[-(\dot t^2 -
{t^\prime}^2) + \frac{1}{1+\varkappa^2\cos^2\theta}(\dot\theta^2-{\theta^\prime}^2) +
 \frac{\sin^2\theta}{1+\varkappa^2\cos^2\theta}(\dot\varphi^2 - {\varphi^\prime}^2) +\cos^2\theta
(\dot\phi^2 - {\phi^\prime}^2) \Big].
\end{equation}
We chose the following simple ansatz for studying the pulsating and rotating string
in the above mentioned space
\begin{equation} t=
t(\tau), ~~~~~~ \theta =\theta(\tau), ~~~~~~ \varphi =m \sigma, ~~~~~~
\phi = \phi(\tau).
\end{equation}
Instead of solving the equation of motion, we concentrate on the
Virasoro constraint
\begin{equation}
g_{MN}(\partial_{\tau}X^M \partial_{\tau}X^N +\partial_{\sigma}X^M \partial_{\sigma}X^N)=0 ,
\end{equation}
which gives us the following equation for the evolution of $\theta$ as
\begin{equation}
 -\dot t^2 + \frac{1}{1+\varkappa^2\cos^2\theta} \dot\theta^2 + \frac{\sin^2\theta}{1+\varkappa^2\cos^2\theta} m^2 + \cos^2\theta \dot\phi^2 = 0 \label{10}
\end{equation}
The conserved Noether charges are given by
\begin{equation}
\mathcal E = \dot t, ~~~~~~~~~~~ \mathcal J = \cos^2\theta \dot\phi.
\end{equation}
Putting the above in \ref{10} we get the equation
\begin{equation}
\dot\theta^2 = (1+\varkappa^2\cos^2\theta)\mathcal E^2 -m^2 \sin^2\theta - \frac{(1+\varkappa^2\cos^2\theta)\mathcal
J^2}{\cos^2\theta}.\label{11}
\end{equation}
The above equation looks like the equation of motion of a particle in a classical potential. We note that
the effective potential is finite at $\theta = 0$ while it diverges at $\theta = \pi/2$. So in a classical
perspective this is the equation of a particle oscillating between $\theta = 0$ to some extremum value. We
then define the oscillation number, which is an adiabatic invariant for the system, as
\begin{equation}
\mathcal N = \frac{1}{2\pi} \oint d\theta \sqrt{(1+\varkappa^2\cos^2\theta)\mathcal E^2 -m^2 \sin^2\theta -
\frac{(1+\varkappa^2\cos^2\theta)\mathcal J^2}{\cos^2\theta}}.\label{12}
\end{equation}
Putting $\sin\theta = x$ in the above equation (\ref{12}), we get
\begin{equation}
\mathcal N = \frac{1}{\pi} \int_{0}^{\sqrt R} \frac{dx}{1-x^2}
\sqrt{\mathcal E^2(1-x^2)(1+\varkappa^2(1- x^2)) -m^2 x^2(1-x^2) - \mathcal J^2 (1+\varkappa^2(1- x^2))},
\label{13}
\end{equation}
where $R$ is an appropriate root of the polynomial
\begin{equation}
 g(x)= x^4(\varkappa^2\mathcal{E}^2 + m^2) + x^2(-2\varkappa^2\mathcal{E}^2 - m^2 -\mathcal{E}^2 +\varkappa^2\mathcal{J}^2)
 + (1+\varkappa^2)(\mathcal{E}^2- \mathcal{J}^2). \label{14}
\end{equation}
Differentiating (\ref{13}) w.r.t $m$ we can write
\begin{equation}
\frac{\partial \mathcal N}{\partial m} = -\frac{m}{\pi}
\int_{0}^{\sqrt R}\frac{x^2}{\sqrt{(\mathcal E^2(1-x^2)-\mathcal J^2)(1+\varkappa^2(1- x^2)) -m^2 x^2(1-x^2) }}dx.
\end{equation}
Putting $x^2 = z$ we can write the above integral as follows
\begin{equation}
 \frac{\partial \mathcal N}{\partial m} = -\frac{m}{\pi\sqrt{\varkappa^2\mathcal{E}^2 + m^2}}
\int_{0}^{R_{-}}\frac{z~dz}{\sqrt{z(z-R_{+})(z-R_{-})}} \ ,
\end{equation}
where the roots $R_{\pm}$ are the solutions of (\ref{14}) with $z
= x^2$. It can be worthily noted that the bigger root $R_{+}$ is
greater than 1, hence not appropriate here. In this consideration,
the above integral can be written as a combination of Elliptic
integrals as follows,
\begin{equation}
  \frac{\partial \mathcal N}{\partial m} =\frac{2m\sqrt{R_+}}{\pi
\sqrt{\varkappa^2\mathcal{E}^2 + m^2}} \left[\mathbb {E}\Big(\frac{R_-}{R_+}\Big) - \mathbb
{K}\Big(\frac{R_-}{R_+}\Big)\right]
\end{equation}
$\mathbb{E}$ and $\mathbb{K}$ are the known Elliptic integral of first and second kind respectively.
Now to find a relation between Energy and Angular momenta via the oscillation number, we expand the
above equation in in small $\mathcal E$ and $\mathcal J$ limit to get
\begin{eqnarray}
\frac{\partial \mathcal N}{\partial m} &=& \left[\frac{\mathcal
(1+\varkappa^2)J^2} {2m^2} - \frac{3(1+\varkappa^2)(5+\varkappa^2)\mathcal J^4}{16 m^4} + \mathcal O(\mathcal
J^6)\right]  \nonumber\\
&+& \left[-\frac{(1+\varkappa^2)}{2 m^2} + \frac{(9+6\varkappa^2 -3\varkappa^4) \mathcal J^2}
{8m^4} - \frac{15(\varkappa^6-9\varkappa^4-45\varkappa^4-35)\mathcal J^4}{128 m^6} + \mathcal O(\mathcal
J^6)\right] \mathcal E^2 \nonumber\\
&+&
\left[\frac{3(-1+2\varkappa^2+3\varkappa^4)}{16
m^4}+\frac{45(5+3\varkappa^2-\varkappa^4+\varkappa^6)\mathcal J^2} {128m^6} \right.
\nonumber\\ &-& \left. \frac{105(105+140\varkappa^2+30\varkappa^4-4\varkappa^6+\varkappa^8)\mathcal
J^4}{1024 m^8} + \mathcal O(\mathcal J^6)\right] \mathcal E^4
+\mathcal O(\mathcal E^6). \label{16}
 \end{eqnarray}
Integrating the above equation (\ref{16}) with respect to $m$ and
reversing the series to that of $\mathcal{E}$, we get a very long
and complicated expression. But this expression under the limit
$\varkappa \rightarrow 0$ can be shown to reduce to the expression
\begin{equation}
\mathcal E = \sqrt{2m \mathcal G}~~ A(\mathcal J)\left[ 1 -
B(\mathcal J) \frac{\mathcal G}{8m}+ \mathcal O[\mathcal
G^2]\right], \label{17}
\end{equation}
\begin{eqnarray}
{\rm with} ~~~\mathcal G &=& \mathcal N + \frac{\mathcal J^2}{2
m}-\frac{5\mathcal J^4}{16 m^3}+ \mathcal O[\mathcal J^6] \cr &&
\cr A(\mathcal J) &=& \left[1 - \frac{3\mathcal J^2}{4
m^2}+\frac{105\mathcal J^4}{64 m^4}+ \mathcal O[\mathcal
J^6]\right]^{-1/2} \cr && \cr {\rm also,}~~~ B(\mathcal J) &=&
\left[1 - \frac{45\mathcal J^2}{8 m^2}+\frac{1575\mathcal J^4}{64
m^4}+ \mathcal O[\mathcal J^6]\right] A^4(\mathcal J).
\label{18}
\end{eqnarray}
Which is the exact same relation as obtained recently in
\cite{Pradhan:2013sja} for rotating and pulsating strings in the
undeformed $\mathbf{R}\times S^3$. This small $\mathcal J$ limit
is called the short string limit. The above equation presents the
energy-spin relation of the strings in this limit.
\section{Conclusion} In this paper, we have studied rigidly
rotating and oscillating string solutions in the
$\varkappa$-deformed $AdS_3\times S^3$ background. For the rigidly
rotating string, we have restricted the motion to two subspaces of
the $S^3_{\varkappa}$ by reducing along two isometries. We have
found two limits corresponding to the known giant
magnon\cite{Arutynov:2014ota}, \cite{Khouchen:2014kaa} limit and
the new single spike solution of the string moving in this
background. We have found out the most general form of the charges
and derived the dispersion relation among the charges and the
angular separation between the end points of the string. We also
have  explicitly checked that in the limit $\varkappa\rightarrow
0$ limit they reduce to the well known relation presented, for
example, in \cite{Ishizeki:2007we}. Since it is not understood how
the parameter $\varkappa$ deforms the dual CFT, the realization of
the newly found dispersion relations in the CFT are not clear to
us at present. It would further be interesting to generalize the
results presented here to the case of two angular momenta along
the two isometries of the $S^3_{\varkappa}$. Unfortunately, the
analysis appears to be tricky and we hope to present the results
in a future work. Also it would be interesting enough to study
string dynamics in the two limiting backgrounds mentioned in
\cite{Hoare:2014pna}. Since these can be related to other
integrable models, they might give us more options to gain insight
into the dual field theory. Further we study the oscillating
strings in the $S^3_{\varkappa}$ space and derive the energy of
pulsating string, as function of the oscillation number and
conserved angular momentum. We also check that the relation
reduces to the known one in the $\varkappa\rightarrow 0$ limit.
There are other problems which could be pursued further given the
integrable nature of this background and its reductions. It will
be useful to find more general rotating and wound strings in the
full $(AdS_5\times S^5)_{\varkappa}$ of \cite{Arutyunov:2013ega}
as the $AdS$ part contains a singularity for particular value of
$\varkappa$. We hope to report on these issues in future.
\section*{Acknowledgement} We thank the anonymous referee for constructive suggestions.
We would like to thank R. Borsato and Stijn van Tongeren
for making very important comments about our paper and bringing to
our attention \cite{Arutyunov:2013ega} and
\cite{Arutyunov:2014cra}. KLP would like to thank A. Tseytlin for
a correspondence. He would also like to thank CERN PH-TH for
generous hospitality and financial support where a part of this
work was done.

\end{document}